\documentclass[12pt]{article}
\usepackage{amsmath}
\usepackage[english]{babel}
\usepackage{amsthm}
\usepackage{amsfonts}
\usepackage{graphicx}
\usepackage{hyperref}
\usepackage[margin=1.5in]{geometry}
\usepackage[superscript]{cite}
\usepackage{slashed}
\usepackage{comment}
\usepackage{cite}

\theoremstyle{definition}
\newtheorem{definition}{Definition}[section]
\newtheorem{theorem}{Theorem}[section]
\newtheorem{corollary}{Corollary}[theorem]

\theoremstyle{remark}

\usepackage{authblk}

\providecommand{\keywords}[1]
{
\small
\textbf{\textit{Keywords---}} #1
}

\begin{document}
\title{Gateway-like absurdly-benign traversable wormhole solutions}
\author{A. K. Mehta\footnote{Email: abhishek.mehta@students.iiserpune.ac.in}}
\affil{Indian Institute of Science Education and Research, Pune, India}
\maketitle
\abstract{
A class of wormhole solutions is constructed that has restricted polar degrees of freedom to achieve a gateway-like configuration. This compels the use of distribution-valued metrics and connections which further compels the use of neutrix product of distributions, to define distribution-valued curvature, Einstein tensor, and other relevant quantities. The solution demands a spacetime with non-Riemannian effects like non-metricity to be consistent and well-defined, due to the non-associativity of the neutrix product. Finally, the ideal gateway configuration where the negative energy requirement is zero is derived.}\\\\
\newpage
\tableofcontents

\section{Introduction}
Wormholes are a class of solutions in gravity (Einstein or Modified) that connect two distant points in space, allowing for travel in a finite coordinate time of the order of the average human lifespan. Therefore, wormholes, if plausible (and traversable) are an important mode of travel over vast interstellar distances. Wormholes were first conceived as Einstein-Rosen bridges\cite{einstein1935particle}, but they are not traversable due to a singularity present at their throat\cite{Guendelman:2009er}. The first class of traversable wormhole solutions was derived by Ellis\cite{ellis1973ether}, Bronikkov\cite{bronnikov1973scalar}, Cl\'{e}ment\cite{clement1984class}. A subclass of these solutions is called Absurdly Benign Traversable Wormholes (ABTW) \cite{woodward2011making, garattini2020generalized}, first studied by Morris and Thorne\cite{morris1988wormholes}.  However, these subclass of wormhole solutions require a great deal of exotic matter or negative energy in their construction\cite{morris1988wormholes}. Here, ``absurdly benign" means that the wormhole is safe for human travel without suffering damage due to the tidal forces. ABTW also makes a frequent presence in science fiction franchises as `gateways' to different parts of the universe. However, unlike the ABTW in literature, which are always considered to be spherically symmetric, these fictional `gateways' are not spherically symmetric. They are mostly disk-like and behave literally like a `gateway'\footnote{In particular, the \emph{Stargate} franchise.}.

In this paper, taking an inspirational note of the above, we try to develop `gateway-like' wormhole solutions by restricting the polar degrees of freedom of the simplest Morris-Thorne wormhole\footnote{It is worth mentioning that Morris-Thorne wormholes are special cases of the traversable wormholes studied and derived previously. Here, `Morris-Thorne' wormhole is used in the sense of simplest ABTWs.}. The final version is not exactly disk-like but a bulging convex section of a sphere with the throat radius determining its convexity. This restriction, however, forces us to work with discontinuous metrics, for which appropriate formalism is developed to work with them consistently. The metric is treated like a distribution and using the neutrix product defined on the space of distributions, the distribution-valued connections and curvature are defined. Due to the non-associativity of the product, it is more convenient to work with Palatini formalism of gravity and treat the connection as a separate distribution. This allows us to use the product consistently. Eventually, we work out the ideal gateway-like solution which requires no negative energy. We will see that the negative energy matter is replaced by hyperfluids.
\section{Wormhole basics}
The simplest example of a Morris-Thorne wormhole metric is given by\cite{morris1988wormholes}
\begin{align}
ds^2 = -dt^2 + dl^2 + (l^2+r^2_0)(d\theta^2+\sin^2\theta d\phi^2)'
\end{align}
where $l \in (-\infty, \infty)$. The entrance or the throat of the wormhole lies at $l = 0$. $r_0$ is called the throat radius. With this metric one may compute the Einstein tensor to obtain\cite{morris1988wormholes}
\begin{align}
&G_{tt} = -\frac{r^2_0}{(l^2+r^2_0 )^2}\quad G_{ll} = -\frac{r^2_0 }{(l^2+r^2_0)^2}\\
& G_{\theta\theta} = \frac{r^2_0}{l^2+r^2_0}\quad G_{\phi\phi} =\frac{r^2_0}{l^2+r^2_0}\sin^2\theta
\end{align}
Assuming that the Einstein tensor is supported by a stress-energy tensor such that the Einstein equations hold, then the energy density observed by a static observer $u^{\mu} = \delta^{\mu}_t$ is given by
\begin{align}
8\pi\rho = 8\pi T_{\mu\nu}u^{\mu}u^{\nu} = G_{tt}\label{MTV}
\end{align}
Notice that the energy density is negative. For completeness, we compute the total negative energy required
\begin{align}
E = \frac{1}{8\pi}\int d^3x \sqrt{h} G_{tt} = -\frac{\pi r_0 }{2}
\end{align}
Since the wormhole is spherically symmetric, the throat of the wormhole is spherical. This is clearly wasteful because we may not have enough negative energy to build a full spherical wormhole of the desired radius at all times. Therefore, with the expectation of being able to reduce the amount of negative energy required, we will restrict the polar degrees of freedom $(\theta)$ so that the wormhole more closely resembles a gateway.\footnote{Henceforth, the term `stargate' will be used to refer to our gateway-like traversable wormhole solutions.}
\section{Stargate metric}
The polar degrees of freedom are restricted by requiring
\begin{align}
&ds_{+}^2 = -dt^2 + dl^2 + (l^2+r^2_0)(d\theta^2+\sin^2\theta d\phi^2), \theta < \theta_0\\
&ds_{-}^2 = -dt^2 + dl^2 + l^2(d\theta^2+\sin^2\theta d\phi^2), \theta > \theta_0 \label{sgm0}
\end{align}
where $+$ denotes the part of spacetime where $\theta < \theta_0$ and $-$ denotes the part of spacetime where $\theta > \theta_0$ i.e. $\mathcal{M} = \mathcal{M}^{+} \cup \mathcal{M}^{-} \cup \Sigma$. Let the metric on $\mathcal{M}^{\pm}$ be $g^{\pm}$ where $g^{+}$ is the wormhole metric and $g^{-}$ is the flat-metric and $\Sigma$ is the 'discontinuity surface'\footnote{Which we also call the interface.}. Now, we write the full metric $g$ on $\mathcal{M}$ as a distribution using $\Theta$-functions
\begin{align}
g_{\alpha\beta} = \Theta(f)g^{+}_{\alpha\beta}+\Theta(-f)g^{-}_{\alpha\beta}
\end{align}
where $f = \theta_0-\theta$ and $\Sigma$ is given by $f = 0$\footnote{Due to the conical discontinuity at at $\theta = \theta_0$, one may naively expect a conical singularity at the origin. However, a naive computation of the curvature tensor due to the above in metric gravity shows that the only singularities are of the form $\delta(\theta-\theta_0)$ arising due to the discontinuity. Refer to Appendix \ref{cs} for additional details.}. Therefore, instead of the full spherical throat of the Morris-Thorne wormhole (MTW), only an angular section of the throat is accessible. It is now more appropriate to call it an entrance. Now, we will try all possible ways to support this solution through some matter fields. If we consider the derivative of the above, we find that
\begin{align}
\partial_{\gamma}g_{\alpha\beta} = \Theta(f)\partial_{\gamma}g^{+}_{\alpha\beta}+\Theta(-f)\partial_{\gamma}g^{-}_{\alpha\beta} + \delta(f)n_{\gamma}[g_{\alpha\beta}]
\end{align}
where $n_{\gamma} = \partial_{\gamma}f$ and $[g] = g^{+}-g^{-}$. Notice that immediately we run into problems because the curvature tensor $R \sim \partial g\cdot \partial g$ leads to product of distributions. To deal with these products consistently, one has to introduce a well-defined algebra where product among distributions and product among distributions and ordinary\footnote{By ordinary, we mean at least $C^1$ differentiable.} functions are well-defined, such that, it reduces to the usual product for ordinary functions. Such a product exists and is called the neutrix product developed by van der Corput\cite{van1959introduction}, Fisher\cite{fisher1971product} and Mikusi\'{n}ski\cite{MR203392}. The product is represented by $\circ$ and under the $\circ$-product the following product of distributions hold
\begin{align}\label{dlist}
&\Theta(x)^2 = \Theta(x)\circ \Theta(x)= \Theta(x)\notag\\
&\Theta(x)\circ \Theta(-x) = 0\notag\\
&\Theta(-x)\circ \delta(x) = \Theta(x)\circ \delta(x) = \frac{1}{2}\delta(x)\notag\\
& \delta^2(x) = \delta(x)\circ \delta(x) = 0
\end{align}
The above are the only ones relevant to our computation. Throughout the paper all products of distribution-valued quantities are of the neutrix kind. However, it must be noted that these products are not associative\cite{fisher1971product}, since,
$$(\Theta(x)\circ\delta(x))\circ \Theta(-x) = \frac{1}{2}\delta(x)\circ\Theta(-x) = \frac{1}{4}\delta(x)$$
but
$$(\Theta(x)\circ\Theta(-x))\circ \delta(x) = 0$$
Hence, in any computation one must try to avoid exceeding more than two neutrix products. A review of the neutrix products of distributions is given in Appendix \ref{Review}.

\subsection{Metric-compatible connection}
Consider the distribution-valued metric we defined earlier\footnote{Throughout the paper, objects with $+$ superscripts refer to the wormhole and $-$ superscripts refer to the flat-space. }
\begin{align}
g_{\alpha\beta} = \Theta(f)g^{+}_{\alpha\beta}+\Theta(-f)g^{-}_{\alpha\beta}\label{sgm}
\end{align}
If we consider the following distribution-valued connection
\begin{align}
\Gamma^{\gamma}_{\alpha\beta}= \Theta(f)\Gamma^{+\gamma}_{\alpha\beta}+\Theta(-f)\Gamma^{-\gamma}_{\alpha\beta}+\delta(f) \frac{\left(\bar{g}^{-1}\right)^{\gamma\tau}}{2}\left(n_{\alpha}[g_{\tau\beta}]+n_{\beta}[g_{\tau\alpha}]-n_{\tau}[g_{\alpha\beta}]\right)
\end{align}
where $\Gamma^{+}$ being the metric connection for the MTW and $\Gamma^{-}$ being the metric connection for flat-space in radial coordinates and
\begin{align}
\bar{g}_{\alpha\beta} = \frac{g^{+}_{\alpha\beta}+g^{-}_{\alpha\beta}}{2} \quad [g_{\alpha\beta}] = g^{+}_{\alpha\beta}-g^{-}_{\alpha\beta}
\end{align}
Then we find that
\begin{align}
&\nabla_{\mu}g_{\alpha\beta} = 0 \quad \nabla_{\mu}g^{\alpha\beta} = 0
\end{align}
showing that the distribution-valued connection is metric compatible\footnote{See Appendix \ref{CDG}.}. Notice that the connection compatible to the metric is not the Levi-Civita or the metric connection. The derivative of the connection then becomes
\begin{align}
&\partial_{\delta}\Gamma^{\gamma}_{\alpha\beta} = \Theta(f)\partial_{\delta}\Gamma^{+\gamma}_{\alpha\beta}+\Theta(-f)\partial_{\delta}\Gamma^{-\gamma}_{\alpha\beta}+\delta(f)n_{\delta}[\Gamma^{\gamma}_{\alpha\beta}]+\partial_{\delta}(\delta(f)A^{\gamma}_{\alpha\beta})
\end{align}
where
\begin{align}
A^{\gamma}_{\alpha\beta} =\frac{(\bar{g}^{-1})^{\gamma\tau}}{2}\left(n_{\alpha}[g_{\tau\beta}]+n_{\beta}[g_{\tau\alpha}]-n_{\tau}[g_{\alpha\beta}]\right)
\end{align}
and $[\Gamma] = \Gamma^{+}-\Gamma^{-}$. For clarity, we write the non-zero components of $[\Gamma]$, $[g]$
\begin{align}
&[\Gamma^{\theta}_{l\theta}] = [\Gamma^{\phi}_{l\phi}] = \frac{l}{l^2+r^2_0}-\frac{1}{l}\notag\\
&[g_{\theta\theta}] = r^2_0 \quad [g_{\phi\phi}] = r^2_0\sin^2\theta\label{disc}
\end{align}
Therefore, we have
\begin{align}
R^{\mu}_{~\nu\rho\sigma} = \Theta(f) R^{+\mu}_{~~~\nu\rho\sigma} + \Theta(-f) R^{-\mu}_{~~~\nu\rho\sigma}+\delta(f)n_{[\rho}[\Gamma^{\mu}_{\sigma]\nu}]+\nabla_{\tau}(\delta^{\tau}_{\rho}\delta(f) A^{\mu}_{\nu\sigma}-\delta^{\tau}_{\sigma}\delta(f)A^{\mu}_{\nu\rho})\label{set1}
\end{align}
where we have implicitly assumed a neutrix product and used (\ref{dlist}) in the above\footnote{It is worth pointing out that in the limit of $\theta_0 = \pi$, the curvature tensor reads $R^{\mu}_{~\nu\rho\sigma}\big|_{\theta_0 = \pi} = \Theta(\pi-\theta) R^{+\mu}_{\nu\rho\sigma} -\delta(\pi-\theta)\left[-n_{[\rho}[\Gamma^{\mu}_{\sigma]\nu}]+\frac{4r^2_0l}{(2l^2+r^2_0)^2}(\delta^{l}_{\rho}\delta^{\theta}_{\sigma}-\delta^{l}_{\sigma}\delta^{\theta}_{\rho})\delta^{\mu}_{\theta}\delta^{\theta}_{\nu}\right]$ i.e. the MTW behaviour is reproduced up to $\delta$-function term. However, if we consider the Riemann curvature density of weight 1 defined by $\tilde{R}^{\mu}_{~\nu\rho\sigma} = \sqrt{g}R^{\mu}_{~\nu\rho\sigma}$, then we find that $\tilde{R}^{\mu}_{\nu\rho\sigma}\big|_{\theta_0 = \pi} = \tilde{R}^{+\mu}_{\nu\rho\sigma}$ i.e. reproducing the MTW behaviour completely. Therefore, this $\delta$-function term is just an artifact of the distribution-valued curvature and is not a real singularity.}. Since, we are only going up to a double product of the distribution-valued connection the Riemann tensor derived above is well-defined. However, to compute the Einstein tensor $G_{\mu\nu}$ one has to contract the above with $g_{\mu\nu}\circ g^{\nu\sigma}$ and due to this, the third term in (\ref{set1}) will lead to a product of distribution of the form
\begin{align}
\Theta(x)\circ\delta(x)\circ \Theta(-x)
\end{align}
which we have shown to be inconsistent before. Hence, the Einstein tensor is ill-defined due to the non-associativity of the product. This means that the stargate metric cannot be supported with a metric-compatible connection.

\subsection{Non-metricity}\label{nms}
In the previous section, we saw that due to the non-associativity of the neutrix product, we were unable to consistently define the Einstein tensor. Due to this non-associativity, it is, therefore, convenient to define the metric and the connection as independent distributions which is precisely the Palatini formalism of general relativity. Henceforth, we abandon the metricity condition altogether and now consider the following connection
\begin{align}
\Gamma^{\gamma}_{\alpha\beta}= \Theta(f)\Gamma^{+\gamma}_{\alpha\beta}+\Theta(-f)\Gamma^{-\gamma}_{\alpha\beta}\label{nmc}
\end{align}
where we find that
\begin{align}
\nabla_{\mu}g_{\alpha\beta} = \delta(f) n_{\mu}[g_{\alpha\beta}]
\end{align}
which is clearly not metric compatible. The Riemann tensor then reads\footnote{Here as well, in the limit of $\theta_0 = \pi$, the curvature tensor reproduces the MTW behaviour up to a $\delta$-function term $R^{\mu}_{~\nu\rho\sigma}\big|_{\theta_0 = \pi} = \Theta(\pi-\theta) R^{+\mu}_{\nu\rho\sigma} +\delta(\pi-\theta)n_{[\rho}[\Gamma^{\mu}_{\sigma]\nu}]$. However, one can again see that $\tilde{R}^{\mu}_{\nu\rho\sigma}\big|_{\theta_0 = \pi} = \tilde{R}^{+\mu}_{\nu\rho\sigma}$, hence, reproducing the MTW behaviour completely.}
\begin{align}
R^{\mu}_{~\nu\rho\sigma} = \Theta(f) R^{+\mu}_{~~~\nu\rho\sigma} + \Theta(-f) R^{-\mu}_{~~~\nu\rho\sigma}+\delta(f)n_{[\rho}[\Gamma^{\mu}_{\sigma]\nu}] \label{R2}
\end{align}

\subsubsection{Energy requirement}
Consider the distribution-valued Riemann curvature tensor (\ref{R2}) where we contract $\mu, \rho$ index to obtain a distribution-valued Ricci tensor
\begin{align}
R_{\nu\sigma} = \Theta(f) R^{+}_{\nu\sigma}+ \Theta(-f)R^{-}_{\nu\sigma} + \delta(f)\left(n_{\mu}[\Gamma^{\mu}_{\sigma\nu}]-n_{\sigma}[\Gamma^{\mu}_{\mu\nu}] \right)\label{Ricd}
\end{align}
Now, to obtain the Ricci scalar, we have to evaluate the product of distributions $g^{\nu\sigma}\circ R_{\nu\sigma}$. This product exists and is given by
\begin{align}
R \equiv g^{\nu\sigma}\circ R_{\nu\sigma} = \Theta(f) R^{+}+ \Theta(-f) R^{-}\label{scalar}
\end{align}
The $\delta$-function part vanishes as it computes to zero, see (\ref{sst}). Now, to compute the Einstein tensor, we require the product $g_{\nu\sigma}\circ R$ more explicitly $g_{\nu\sigma}\circ g^{\alpha\beta} \circ R_{\alpha\beta}$ . This is a triple neutrix product. Since the product is non-associative, a triple product may not have a consistent answer. However, the Ricci scalar (\ref{scalar}) only has the $\Theta$-functions, over which the neutrix products are always associative. Therefore, in this case
\begin{align}
(g_{\nu\sigma}\circ g^{\alpha\beta}) \circ R_{\alpha\beta} = g_{\nu\sigma}\circ (g^{\alpha\beta} \circ R_{\alpha\beta}) = g^{\alpha\beta} \circ (g_{\nu\sigma}\circ R_{\alpha\beta}) = g_{\nu\sigma}\circ R
\end{align}
which means that $g_{\nu\sigma}\circ R$ exists and hence, the distribution-valued Einstein tensor exists as well\footnote{Notice how the vanishing of the $\delta$-function leads to an associative triple neutrix product allowing for an unambiguous computation of the Einstein tensor. This is the reason that makes the non-metricity setting preferable to the metric compatible case.}.
\begin{align}
G_{\nu\sigma} = \Theta(f) G^{+}_{\nu\sigma}+\Theta(-f) G^{-}_{\nu\sigma}+\delta(f)\left(n_{\mu}[\Gamma^{\mu}_{\sigma\nu}]-n_{(\sigma}[\Gamma^{\mu}_{\nu)\mu}] \right)\label{ET}
\end{align}
where
\begin{align}
G_{\mu\nu} = R_{(\mu, \nu)}-\frac{1}{2}g_{\mu\nu} R
\end{align}
Assuming that the above is sourced by the stress-energy tensor $T_{\mu\nu}$, the energy density observed by a static observer $u^{\mu} = \delta^{\mu}_t$ is given by
\begin{align}
8\pi\rho = 8\pi T_{\mu\nu}u^{\mu}u^{\nu} = G_{tt}
\end{align}
Since, $G_{tt}$ only has $\theta$-function terms, any product with other $\theta$-function based distribution will be associative. It is easy to see that
\begin{align}
\sqrt{h} = \Theta(f) \sqrt{h^{+}}+\Theta(-f) \sqrt{h^{-}}
\end{align}
where $h_{\mu\nu}$ is the metric on the constant time hypersurface. Therefore, the total energy is given by
\begin{align}
E = \frac{1}{8\pi}\int d^3x \sqrt{h}\cdot G_{tt} = \frac{1}{8\pi}\int d^3 x \Theta(f) \sqrt{h^{+}} G^{+}_{tt} = -\frac{r_0}{8}\int_{\theta < \theta_0} d\Omega^2_{S^2} = -\frac{\pi r_0 }{2}\sin^2\frac{\theta_0}{2}
\end{align}
since, $G^{-}_{tt} = 0$. Notice that the negative energy requirement is a significant fraction compared to the simplest Morris-Thorne wormhole. It seems some significant and promising progress has been made.
\subsubsection{Interface stress-energy tensor}
However, if we explicitly compute the $\delta$-function component of the Einstein tensor (\ref{ET}), we find that for our case
\begin{align}
G_{\nu\sigma}\big |_{\delta} = n_{\mu}[\Gamma^{\mu}_{\sigma\nu}]-n_{(\sigma}[\Gamma^{\mu}_{\nu)\mu}] = 0\label{sst}
\end{align}
using (\ref{disc}). The Palatini spacetime is also accompanied by a hypermomentum tensor density or a superpotential
\begin{align}
-\nabla_{\lambda}(\sqrt{-g}g^{\mu\nu})+\nabla_{\sigma}(\sqrt{-g}g^{\sigma(\mu})\delta^{\nu)}_{\lambda} = -\Delta^{~\mu\nu}_{\lambda}\label{superpot}
\end{align}
This superpotential is the consequence of the variation due to the connection namely
\begin{align}
\Delta^{~\mu\nu}_{\lambda} = -\frac{\delta S}{\delta \Gamma^{\lambda}_{\mu\nu}}
\end{align}
More explicitly, for our case
\begin{align}
\Delta^{~\mu\nu}_{\lambda} = \delta(f)\delta^{\theta}_{\lambda}(-\delta^{\mu}_{t}\delta^{\nu}_t+\delta^{\mu}_{l}\delta^{\nu}_{l})r^2_0\sin\theta_0\label{hfluid}
\end{align}
This hypermomentum can be sourced using some hyperfluid $\Delta^{f\mu\nu}_{\lambda}$ which are a special kind of fluid that have intrinsic spin and other microstructures\cite{iosifidis2020cosmological, iosifidis2020non, puetzfeld2008probing}. The hyperfluids obey the following continuity equation in absence of torsion\cite{iosifidis2020cosmological}
\begin{align}
\tilde{\nabla}^{\mu}\mathcal{T}_{\mu\nu}+\nabla_{\rho}\nabla_{\sigma}\Delta_{\nu}^{~\rho\sigma}-R^{\lambda}_{~\rho\sigma\nu}\Delta_{\lambda}^{~\rho\sigma} = 0\label{eoq}
\end{align}
where $\tilde{\nabla}$ is the covariant derivative with respect to the metric connection and $\mathcal{T}_{\mu\nu}$ is the stress-energy tensor density. However, it turns out that the continuity equation is not satisfied in this configuration, see \ref{nosat}. Therefore, such a configuration, promising as they may be, is unphysical. Also, there are no geodesics that pass through the interface, see Section \ref{tf}. Hence, the interface in this case is actually non-traversable.

\section{The ideal stargate}\label{dream}
An ideal stargate configuration should have the minimum negative energy requirement\footnote{Ideally, none.}, have a traversable interface and the matter supporting the configuration should be physical. Despite the lack of success in the previous section, there is still hope in abandoning metric-compatibility, as one has many suitable connection choices that can achieve the desired stargate configuration. We find that this is most ideally achieved by the choice
\begin{align}
\Gamma^{\gamma}_{\alpha\beta} = \Gamma^{-\gamma}_{\alpha\beta}~~ (\equiv \text{flat-connection})\label{cc}
\end{align}
with the metric still being (\ref{sgm}). If we look at the equations of motion due to the Palatini variation for a torsionless connection
\begin{align}
&R_{(\mu, \nu)}-\frac{1}{2}R g_{\mu\nu} = T_{\mu\nu}\notag\\
&-\nabla_{\lambda}(\sqrt{-g}g^{\mu\nu})+\nabla_{\sigma}(\sqrt{-g}g^{\sigma(\mu})\delta^{\nu)}_{\lambda} = -\Delta^{\mu\nu}_{\lambda} \label{meom}
\end{align}
then, due to the connection (\ref{cc}) and the metric (\ref{sgm}), we have\footnote{For the case of $\theta_0 = 0$, we obtain the usual flat-space while for $\theta_0 = \pi$, we obtain the full spherical wormhole completely supported by the hyperfluids. Refer to Appendix \ref{limbeh} for additional details.}
\begin{align}
&T_{\mu\nu} = 0 \label{seT}\\
&\Delta^{~\mu\nu}_{\lambda} = \delta(f)\delta^{\theta}_{\lambda}(-\delta^{\mu}_{t}\delta^{\nu}_t+\delta^{\mu}_{l}\delta^{\nu}_{l})r^2_0\sin\theta_0 + \Theta(f)\chi^{~\mu\nu}_{\lambda}\label{hypermomex}\\
&\chi^{~\mu\nu}_{\lambda} =2\frac{ r^2_0}{l}\sin\theta[\delta^{l}_{\lambda}(-\delta^{\mu}_{t}\delta^{\nu}_t+\delta^{\mu}_{l}\delta^{\nu}_{l})-\delta^{(\nu}_l\delta^{\mu)}_{\lambda}]
\end{align}
which can be made to work given the appropriate hyperfluid. And, unlike the case of Section \ref{nms}, the equation of continuity (\ref{eoq}) is trivially satisfied by this configuration (see \ref{ayesat}). This choice means that not only that there is no need for negative energy, the geodesic equation has consistent solution through the interface. And most importantly, there are no tidal forces as there is no curvature i.e. we have a "benign" stargate configuration (see Section \ref{dct?}). If we allow for torsion then
\begin{align}
\Gamma^{\gamma}_{\alpha\beta} = (\Lambda^{-1})^{\gamma}_{~\sigma}\partial_{\beta} \Lambda^{\sigma}_{~\alpha}~~ (\equiv \text{inertial-connection})
\end{align}
where $\Lambda \in GL(4, \mathbb{R})$ will also give the zero energy configuration while still satisfying the equation of continuity. However, the geodesic deviation equation will get contributions due to the torsion and therefore, the stargate solutions may not be ``benign".
\subsection{Hyperfluid source}\label{hypfluid}
The hyperfluids are usually thought to be ordinary fluids with some microstructure to them. For instance, a fluid made of fermions will have a spin microstructure to it. The macroscopic properties of the fluid source the stress-energy tensor $T_{\mu\nu}$ while the microstructures of the fluid source the $\Delta^{\mu}_{\nu\lambda}$. One can also imagine different kinds of fluids with very exotic microstructures depending on their fundamental constituents, for instance, strings, quarks, etc. However, the hyperfluid required in our ideal stargate configuration should have a zero stress-energy tensor and a non-zero hypermomentum. This is only possible if somehow the hyperfluid of our interest doesn't couple to the metric and only couples to the connection. It is very difficult to come up with such an action that can allow for that and even more difficult to motivate such an action. As an example, however, consider an additional interaction of the form\footnote{This is very reminiscent of Born-Infeld gravity\cite{Olmo:2015wwa}}
\begin{align}
S_{int} = \eta\int d^4x \sqrt{\det (m_{\mu\nu}+\alpha R_{\mu\nu})} + \beta \int d^4x \sqrt{m} + \gamma \int d^4x \sqrt{\det(R_{\mu\nu})}\label{hypfluact}
\end{align}
where $m_{\mu\nu}$ is simply an auxiliary tensor field and $\eta, \beta, \gamma$ are simply coupling constants. The equation of motion with with respect to $m_{\mu\nu}$ and a torsionless $\Gamma^{\mu}_{\nu\lambda}$ is given by
\begin{align}
&\left(\eta\frac{\sqrt{\det (m_{\mu\nu}+\alpha R_{\mu\nu})} }{\sqrt{m}}+\beta\right)m_{\mu\nu} = -\alpha\beta R_{\mu\nu}\\
&\eta\Pi_{\lambda}^{\mu\nu}(m, \Gamma) + \gamma \Pi_{\lambda}^{\mu\nu}(0, \Gamma) = -\Delta^{\mu\nu}_{\lambda}\\
&\Pi_{\lambda}^{\mu\nu}(m, \Gamma) = -\nabla_{\lambda}\left[\sqrt{\det (m_{\rho\sigma}+\alpha R_{\rho\sigma})}[(m+\alpha R)^{-1}]^{\mu\nu}\right]\notag\\&+\nabla_{\sigma}\left[\sqrt{\det (m_{\rho\sigma}+\alpha R_{\rho\sigma})}[(m+\alpha R)^{-1}]^{(\mu|\sigma|}\right]\delta^{\nu)}_{\lambda}
\end{align}
where $\Delta^{\mu\nu}_{\lambda}$ is given by (\ref{meom}). It is not very difficult to see that for $\beta = -\eta, \gamma = 0$, $m_{\mu\nu} = g_{\mu\nu}$ solves the above equation of motion i.e. $g_{\mu\nu}$ being the stargate solution. What we see in (\ref{hypfluact}) is an interaction that only involves the connection and it helps source a non-zero hypermomentum with zero stress-energy tensor. However, it seems quite difficult to motivate such interactions from known phenomena or theories of gravity. Although, it might be possible to eliminate or constrain such kind of interactions by considering no-ghost theorems and locality.
\section{Discussion and Conclusion}
We started with a metric-compatible connection and tried to accommodate what is essentially a discontinuous metric. After a series of trials and errors, we finally settled with a non-metricity spacetime with a flat-connection to achieve a `benign' stargate configuration. The discontinuous metric and non-metricity are non-Riemannian effects, therefore, it is not surprising that the stargate solution is eventually supported by hyperfluids as they have all the necessary microstructures to source such non-Riemannian degrees of freedom\cite{iosifidis2020non, puetzfeld2008probing}. One can in principle develop a class of stargate configurations where the need for hyperfluids may be further reduced to the vicinity of the `entrance'. However, that is a repetitive and perhaps, not-so-illuminating exercise. We also saw that it was a bad idea to support discontinuous metrics with a distribution-valued connection\footnote{Or more simply discontinuous connections} as in presence of such connections, geodesics (metric or affine) through the interface are absent. Trivial accessibility to the stargate is important for its utility and ease of use. Therefore, it seems any stargate solution must necessarily have a continuous connection to ensure that. Also, the matter fields required must at least satisfy the continuity equation. The hyperfluid sources supporting a zero energy stargate are already peculiar enough because such sources must only couple to the connection and not the metric. We made attempts to exemplify such a scenario using (\ref{hypfluact}) but it lacked any fundamental or phenomenological motivation. However, one can check if such interactions are free of ghosts or are local to ascertain their sensibility. Such examples, if sensible, can source hyperfluids for ideal stargates or wormholes. It is possible to source such kinds of hyperfluids via the spacetime itself. The existence of quantum gravity necessitates an underlying spacetime microstructure and vice-versa. Such microstructures would provide some corrections to the usual gravitational action perhaps of the kind we worked with in (\ref{hypfluact}). Maybe this is the phenomenological motivation we have been missing. This line of thought, however, is currently beyond the scope of this paper and hence, will be seriously considered in our future studies.\\

Having alluded to a possibility of operating wormholes without negative energy, one now, finally, has to confront the problem of wormhole construction. Even if we have all the exotic matter needed, it is not clear how would one use the exotic matter to assemble a wormhole. Also, every step of the process of its assembly must be compatible with general relativity, therefore, what we are finally looking at is a time-dependent solution from its construction and deconstruction after its use. Also, such a wormhole assembly requires us to manipulate topology which is forbidden by classical general relativity as it is fundamentally Riemannian in nature\cite{morris1988wormholes}. That is where the non-Riemannian nature of our solution and the sources that support it may prove to be useful. This is because one can in principle consider the stargate solution as an intermediate step in the full spherical wormhole assembly. How this step can be initiated from scratch is again a topic that is currently beyond the scope of this paper and, therefore, will be seriously considered in our future studies.
\section*{Acknowledgement}
This work is dedicated to the people and culture of India, that is Bharat, for steadily supporting research in basic sciences for thousands of years and continues to do so even to this day. I would like to thank José Senovilla (UPV/EHU) and Damianos Iosifidis (AUTH) for their prompt clarifications and excellent explanations of all things GR. I would also like to acknowledge the support of the CSIR-UGC (JRF) fellowship (09/936(0212)/2019-EMR-I).
\section*{Appendix}
\appendix
\section{Review of product of distributions}\label{Review}
This is a review of product of distributions developed by van der Corput\cite{van1959introduction}, Fisher\cite{fisher1971product}. Let $\rho$ be a fixed infinitely differentiable function such that
\begin{align}
&\rho(x) = 0 \forall |x| \geq 1\\
& \rho(x) \geq 0\\
&\rho(x) = \rho(-x)\\
&\int^{1}_{-1}\rho(x) dx = 1
\end{align}
Now, define a function $\delta_n(x)$ by
\begin{align}
\delta_n(x) = n\rho(nx) \quad n = 1, 2, \cdots
\end{align}
$\{\delta_n\}$ is now a sequence of infinitely differentiable functions converging to the delta function $\delta(x)$ as $n \to \infty$. Consider an arbitrary distribution $f$ and define
\begin{align}
f_n(x) = f * \delta_n = \int^{1/n}_{-1/n}f(x-t)\delta_n(t) dt
\end{align}
Then $\{f_n\}$ is a sequence of infinitely differentiable functions converging to the distribution $f$.
\begin{definition}\label{proddef1}
Let $f$ and $g$ be arbitrary distributions and let
$$f_n = f*\delta_n \quad g_n = g * \delta_n$$
Then the product of $f$ and $g$ exists and is equal to the distribution $h = f\cdot g$ on the opern interval $(a, b)$, where $-\infty \leq a < b \leq \infty$, if and only if $\{f_n\cdot g_n\}$ is a regular sequence converging to $h$ on an open interval $(a, b)$.
\end{definition}

\begin{definition}\cite{van1959introduction}
Let $f$ and $g$ be arbitrary distributions and let
$$f_n = f*\delta_n \quad g_n = g * \delta_n$$
Then the neutrix product of $f$ and $g$ exists and is equal to the distribution $h = f\circ g$ on the opern interval $(a, b)$, where $-\infty \leq a < b \leq \infty$, if and only if
\begin{align}
\underset{n\to \infty}{\operatorname{N-lim}}~f_n g_n = h
\end{align}
where $\operatorname{N-lim}$ is called the neutrix limit, $N$ is the neutrix, having the domain $N' = \{1, 2,\cdots, n, \cdots\}$ and the range $N''$ the real numbers, with negligible functions finite linear sums of the functions $$n^{\lambda}\ln^{r-1}n, \quad \ln^r n$$
for $\lambda \geq 0$ and $r = 1, 2,, \cdots$ and all functions which converge to zero in the normal sense as $n$ tends to infinity.
\end{definition}
Notice that if
\begin{align}
\lim_{n\to \infty}f_n g_n = h
\end{align}
then $f \circ g$ reduce to $f \cdot g$, so in a way neutrix product is a generalization of product defined in Definition \ref{proddef1}. Hence, one can state the following theorem due to Fisher-Lin-Zhi\cite{fisher1992product}\\\\
``Let $f$ and $g$ be arbitrary distributions and let $f\cdot g$ exists and is equal to $h$, then the neutrix product $f \circ g$ exists and is equal to $h$.\label{FisherLinZhi}''\\\\
Using the above definition for the product of distributions, one can prove several theorems involving product of distributions. For instance, the theorem by Fisher\cite{fisher1971product}\\\\
``Let $x^{\lambda}_{+} \equiv x^{\lambda}\Theta(x)$ and $x^{\lambda}_{-} \equiv |x|^{\lambda}\Theta(-x)$ then the product $x^{\lambda}_{+}\cdot x^{\mu}_{-}$ exists and is given by
$$x^{\lambda}_{+}\cdot x^{\mu}_{-} = 0$$
provided $\lambda, \mu, \lambda+\mu > -1$. One can also show
$$x^r_{+}\cdot \delta^{(r)}(x) = \frac{(-1)^r}{2}r!\delta(x) \quad r = 0, 1, 2\cdots$$
where $\delta^{(r)}$ is the $r$-th derivative of the delta function.\label{Fisher}''\\\\
We also quote some useful theorems by Mikusi\'{n}ski\cite{MR203392} and Koh-Kuan\cite{koh1992distributions}, respectively\\\\
``$\delta(x)\cdot x^{-1} = -\frac{1}{2}\delta'(x)$\label{Mikusinski}''\\
``Under a neutrix product
\begin{align}
&\delta^{2(l+1)}(x) = 0 \notag\\
&\delta^{2l+1}(x) = C_l\delta^{(2l)}(x)\notag
\end{align}
for $l = 0, 1, 2, 3, \cdots$ where $C_l = \frac{1}{2^{2l}l! (2l+1)^{l+1/2}\pi^l}.$\label{KK}''\\\\
Using the above theorems, one can easily show
\begin{corollary}
In the theorem by Fisher, by setting $\lambda = \mu = 0$ we obtain
$$\Theta(x)\cdot \Theta(-x) = 0$$
and by setting $r = 0$ we obtain
$$\Theta(x)\cdot \delta(x) = \frac{1}{2}\delta(x)$$
\end{corollary}
Another product of distributions we require is $\Theta(x)^2 = \Theta(x)\cdot \Theta(x)= \Theta(x)$\cite{fisher1971product} which is by definition of the Heaviside $\theta$-function. Now, we have all the distributions we need with their products well-defined.
\section{Metric-compatible connection for a distribution-valued metric}\label{CDG}
Consider the metric (\ref{sgm})
\begin{align}
g_{\alpha\beta} = \Theta(f)g^{+}_{\alpha\beta}+\Theta(-f)g^{-}_{\alpha\beta}
\end{align}
For the following distribution-valued connection
\begin{align}
\Gamma^{\gamma}_{\alpha\beta}= \Theta(f)\Gamma^{+\gamma}_{\alpha\beta}+\Theta(-f)\Gamma^{-\gamma}_{\alpha\beta}+\delta(f) \frac{\left(\bar{g}^{-1}\right)^{\gamma\tau}}{2}\left(n_{\alpha}[g_{\tau\beta}]+n_{\beta}[g_{\tau\alpha}]-n_{\tau}[g_{\alpha\beta}]\right)\label{mcc}
\end{align}
we have\footnote{For any object $A$ with a discontinuity, we define on the 'discontinuity surface' $\Sigma$ $$\bar{A} = \frac{A^{+}+A^{-}}{2} \quad [A] = A^{+}-A^{-}$$ where $A^{+}$ and $A^{-}$. $[A]$ is the discontinuity in $A$ across $\Sigma$ and $\bar{A}$ is the mean on $\Sigma$.}
\begin{align}
\nabla_{\mu}g_{\nu\lambda} &= \delta(f)n_{\mu}[g_{\nu\lambda}] - g_{\tau\lambda}\circ\delta(f) \frac{\left(\bar{g}^{-1}\right)^{\tau\delta}}{2}\left(n_{\mu}[g_{\delta\nu}]+n_{\nu}[g_{\delta\mu}]-n_{\delta}[g_{\mu\nu}]\right)\notag\\&-g_{\tau\nu}\circ\delta(f) \frac{\left(\bar{g}^{-1}\right)^{\tau\delta}}{2}\left(n_{\mu}[g_{\delta\lambda}]+n_{\lambda}[g_{\delta\mu}]-n_{\delta}[g_{\lambda\mu}]\right)
\end{align}
Since, the connections $\Gamma^{\pm}$ are compatible with the metric $g^{\pm}$, therefore, the $\theta$-functions do not appear in the above. Now, one can show that the $\delta$-function part cancels. Using the product of distributions, listed in (\ref{dlist})
\begin{align}
g_{\mu\nu}\circ\delta(f) = \bar{g}_{\mu\nu} \delta(f)
\end{align}
we have
\begin{align}
\nabla_{\mu}g_{\nu\lambda}= 0
\end{align}
where
\begin{align}
\bar{g}_{\lambda\tau}\left(\bar{g}^{-1}\right)^{\tau\delta} = \delta^{~\delta}_{\lambda}
\end{align}
was also used. Now, it can be checked that
\begin{align}
g^{\alpha\beta} = \Theta(f)(g^{+})^{\alpha\beta}+\Theta(-f)(g^{-})^{\alpha\beta}
\end{align}
is the inverse of (\ref{sgm}). Similarly, we look at
\begin{align}
\nabla_{\mu}g^{\nu\lambda} &= \delta(f)n_{\mu}[g^{\nu\lambda}]+g^{\lambda\tau}\circ\delta(f) \frac{\left(\bar{g}^{-1}\right)^{\nu\delta}}{2}\left(n_{\tau}[g_{\delta\mu}]+n_{\mu}[g_{\delta\tau}]-n_{\delta}[g_{\tau\mu}]\right)\notag\\&+g^{\nu\tau}\circ\delta(f) \frac{\left(\bar{g}^{-1}\right)^{\lambda\delta}}{2}\left(n_{\tau}[g_{\delta\mu}]+n_{\mu}[g_{\delta\tau}]-n_{\delta}[g_{\tau\mu}]\right)
\end{align}
Using the product of distributions,we obtain
\begin{align}
\nabla_{\mu}g^{\nu\lambda} &= \delta(f)n_{\mu}[g^{\nu\lambda}]+\bar{g}^{\lambda\tau}\delta(f) \frac{\left(\bar{g}^{-1}\right)^{\nu\delta}}{2}\left(n_{\tau}[g_{\delta\mu}]+n_{\mu}[g_{\delta\tau}]-n_{\delta}[g_{\tau\mu}]\right)\notag\\&+\bar{g}^{\nu\tau}\delta(f) \frac{\left(\bar{g}^{-1}\right)^{\lambda\delta}}{2}\left(n_{\tau}[g_{\delta\mu}]+n_{\mu}[g_{\delta\tau}]-n_{\delta}[g_{\tau\mu}]\right)
\end{align}
Now, by making use of the observation that
\begin{align}
\bar{g}^{\lambda\tau} \left(\bar{g}^{-1}\right)^{\nu\delta} = -[g^{\lambda\tau}][g^{\nu\delta}]^{-1}
\end{align}
where $[g]^{-1}$ is the inverse of $[g]$, we can cancel the $\delta$-function contribution and obtain
\begin{align}
\nabla_{\mu}g^{\nu\lambda} = 0
\end{align}
Showing that (\ref{mcc}) is a metric-compatible connection.
\section{Geodesics through the interface for distribution-valued connections}\label{tf}
To access the stargate, the interface must be traversable. This means that there must exist geodesics through the interface. Consider the geodesic equation for a metric connection given by
\begin{align}
\frac{d^2x^{\mu}}{d\tau^2} + \tilde{\Gamma}^{\mu}_{\alpha\beta}\dot{x}^{\alpha}\dot{x}^{\beta} = 0\label{prop}
\end{align}
where for our case the metric connection is given by
\begin{align}
\tilde{\Gamma}^{\gamma}_{\alpha\beta}= \Theta(f)\Gamma^{+\gamma}_{\alpha\beta}+\Theta(-f)\Gamma^{-\gamma}_{\alpha\beta}+\delta(f) \frac{\bar{g}^{\gamma\tau}}{2}\left(n_{\alpha}[g_{\tau\beta}]+n_{\beta}[g_{\tau\alpha}]-n_{\tau}[g_{\alpha\beta}]\right)
\end{align}
Since, we are looking for a path through the interface, therefore, we consider $\dot{\phi} = 0$ i.e. $\phi = \Phi$ which are constants. Then, we have
\begin{align}
&\ddot{t} = 0\\
&\ddot{l}-l\dot{\theta}^2 = 0\\
&\ddot{\theta}+\left(\Theta(f)\frac{l}{l^2+r^2_0}+\Theta(-f)\frac{1}{l}\right)\dot{\theta}\dot{l}+\frac{1}{2}\delta(f)\dot{\theta}^2\left(\frac{r^2_0}{l^2+r^2_0}+\frac{r^2_0}{l^2}\right) = 0\label{geoeq}
\end{align}
Due to the triple distribution product in the second term of (\ref{geoeq}), there are no consistent non-trivial solutions to the above. Therefore, there are no metric geodesics that cross the interface. Since all distribution-valued connections will also face the same issue, therefore, there will be no affine or metric geodesics through the interface for any distribution-valued connections.
\section{Tidal forces in the ideal stargate}\label{dct?}
Since, the ideal stargate configuration arises for a uniform connection, the affine geodesic equation has consistent solutions through the interface. However, as the metric is still distribution-valued, there are no consistent metric geodesics that go through the interface. Now, consider the family of affine geodesics parametrized by $\tau$. The equation for geodesic deviation is then given by
\begin{align}
\frac{D^2\xi^{\mu}}{D\tau^2}-R^{\mu}_{~\nu\rho\sigma} u^{\nu}u^{\rho} \xi^{\sigma} = 0
\end{align}
where $D/D\tau = u^{\mu}\nabla_{\mu}$ and $\xi^{\mu}$ is the deviation vector. Since, we are only interested in the spatial deviations of the geodesics, we introduce the following projector
\begin{align}
h_{\mu}^{~\nu} = \delta_{\mu}^{~\nu}-\frac{u_{\mu} u^{\nu}}{u^2}
\end{align}
where the indices are raised and lowered using (\ref{sgm}). And then define
\begin{align}
\eta^{\nu} \equiv \xi^{\mu}h_{\mu}^{~\nu} = \xi^{\nu}-u^{\nu}\frac{\xi^{\mu} u_{\mu}}{u^2}\label{defproj}
\end{align}
where we have used $Du^{\mu}/D\tau = 0$ as $u^{\mu} = dx^{\mu}/d\tau$ is the tangent to the affine geodesic. Using the above, we find that
\begin{align}
\frac{D\xi^{\nu}}{D\tau} = \frac{D\eta^{\nu}}{D\tau} +u^{\nu}\frac{D}{D\tau}\left( \frac{\xi^{\mu} u_{\mu}}{u^2}\right)\label{intstep}
\end{align}
Now, we introduce the tetrads such that
\begin{align}
e^0_{\mu} = \frac{u_{\mu}}{|u|} \quad e^{a}_{\mu}e^{b}_{\nu}\delta_{ab} = -h_{\mu\nu}
\end{align}
so that we can write (\ref{intstep}) as
\begin{align}
e^{a}_{\nu}\frac{D\xi^{\nu}}{D\tau} = e^{a}_{\nu}\frac{D\eta^{\nu}}{D\tau}
\end{align}
where we have made use of $e^{a}_{\nu} u^{\nu} = e^{a}_{\nu} e^0_{\mu} g^{\mu\nu} |u| = 0$. Using the Liebniz rule, we find that
\begin{align}
\frac{D\xi^a}{D\tau}-\xi^{\nu}\frac{D e_{\nu}^a}{D\tau} = \frac{D\eta^a}{D\tau}-\eta^{\nu}\frac{D e_{\nu}^a}{D\tau}
\end{align}
We make use of (\ref{defproj}), in the above to obtain
\begin{align}
\frac{D\xi^a}{D\tau} = \frac{D\eta^a}{D\tau}+u^{\nu}\left( \frac{\xi^{\mu} u_{\mu}}{u^2}\right)\frac{D e_{\nu}^a}{D\tau}
\end{align}
Using the identity
\begin{align}
0 = \frac{D}{D\tau}(e^{a}_{\nu} u^{\nu}) = u^{\nu}\frac{De^{a}_{\nu}}{D\tau}
\end{align}
we finally have
\begin{align}
\frac{D\xi^a}{D\tau} = \frac{D\eta^a}{D\tau}
\end{align}
which may be used to rewrite the geodesic deviation equation as
\begin{align}
\frac{D^2\eta^a}{D\tau^2}+K^{a}_b\eta^b = 0 \quad K^a_b = -R^{\mu}_{~\nu\rho\sigma} e^{a}_{\mu}v^{\nu}v^{\rho} e^{\sigma}_b
\end{align}
Since, for a flat-connection, like in the ideal stargate configuration, $K^a_b = 0$, therefore, we simply have no tidal forces present i.e.
\begin{align}
\frac{D^2\eta^a}{D\tau^2} = 0
\end{align}
showing that the ideal stargate configuration is also ``absurdly benign".
\section{Equation of Continuity}
\subsection{Derivation}
This is a much simpler version of the derivation given by Iosifidis\cite{iosifidis2020cosmological}. Consider a matter action $S_m$ that is general coordinate invariant. Now, consider an arbitrary translation of the coordinates $x^{\mu} \to x^{\mu}+\xi^{\mu}$, we require
\begin{align}
\delta_{\xi}S_m = \int d^4 x \left[\frac{\delta \mathcal{L}}{\delta g_{\mu\nu}}\delta_{\xi}g_{\mu\nu}+\frac{\delta \mathcal{L}}{\delta \Gamma^{\lambda}_{\mu\nu}}\delta_{\xi}\Gamma^{\lambda}_{\mu\nu}\right] = 0\label{var}
\end{align}
Assuming that the connection is torsionless, one can show that
\begin{align}
&\delta_{\xi}g_{\mu\nu} = -2\tilde{\nabla}_{(\mu}\xi_{\nu)}\\
&\delta_{\xi}\Gamma^{\lambda}_{\mu\nu} = \xi^{\alpha}R^{\lambda}_{~(\mu\nu)\alpha}-\nabla_{(\nu}\nabla_{\mu)}\xi^{\lambda}
\end{align}
where $\tilde{\nabla}$ is the covariant derivative corresponding to the metric connection. Using this in (\ref{var}), we get
\begin{align}
\delta_{\xi}S_m = \int d^4 x \left[-2\mathcal{T}^{\mu\nu} \tilde{\nabla}_{\mu}\xi_{\nu}-\Delta_{\lambda}^{~\mu\nu}( \xi^{\alpha}R^{\lambda}_{~\mu\nu\alpha}-\nabla_{\nu}\nabla_{\mu}\xi^{\lambda})\right]
\end{align}
where we have made use of the following definitions
\begin{align}
&\mathcal{T}^{\mu\nu} := \frac{\delta \mathcal{L}}{\delta g_{\mu\nu}}\\
&\Delta_{\lambda}^{~\mu\nu} := -\frac{\delta \mathcal{L}}{\delta \Gamma^{\lambda}_{\mu\nu}}
\end{align}
where $\mathcal{T}$ is the stress-energy density and $\Delta^{\lambda}_{~\mu\nu}$ is a hypermomentum density. After integration by parts, we obtain
\begin{align}
\tilde{\nabla}^{\mu}\mathcal{T}_{\mu\nu}+\nabla_{\rho}\nabla_{\sigma}\Delta_{\nu}^{~\rho\sigma}-R^{\lambda}_{~\rho\sigma\nu}\Delta_{\lambda}^{~\rho\sigma} = 0
\end{align}
\subsection{Unphysical stargate configuration: $T_{\mu\nu} \neq 0, \Delta_{\mu}^{~\nu\lambda} \neq 0$}\label{nosat}
For (\ref{superpot}), it can be shown that the following identity holds \cite{iosifidis2020cosmological}
\begin{align}
\hat{\nabla}_{\mu}\Delta_{\lambda}^{~\mu\nu} = -\sqrt{-g}g^{\nu\mu}(\check{R}_{\mu\lambda}+R_{\lambda\mu})\label{identity}
\end{align}
where
\begin{align}
\check{R}^{\lambda}_{~\kappa} = R^{\lambda}_{~\mu\nu\kappa}g^{\mu\nu} \quad \hat{\nabla}_{\mu} = 2 S_{\mu} - \nabla_{\mu}
\end{align}
However, for our case the connection (\ref{nmc}) leads to
\begin{align}
\nabla_{\mu}\Delta_{\lambda}^{~\mu\nu} = 0
\end{align}
Hence, the equation of continuity simply reduces to
\begin{align}
\tilde{\nabla}^{\mu}\mathcal{T}_{\mu\nu}-R^{\lambda}_{~\rho\sigma\nu}\Delta_{\lambda}^{~\rho\sigma} = 0\label{eoq2}
\end{align}
The metric connection corresponding to (\ref{sgm}) is simply
\begin{align}
\tilde{\Gamma}^{\gamma}_{\alpha\beta}= \Theta(f)\Gamma^{+\gamma}_{\alpha\beta}+\Theta(-f)\Gamma^{-\gamma}_{\alpha\beta}+\delta(f) \frac{\bar{g}^{\gamma\tau}}{2}\left(n_{\alpha}[g_{\tau\beta}]+n_{\beta}[g_{\tau\alpha}]-n_{\tau}[g_{\alpha\beta}]\right)
\end{align}
Using this in (\ref{eoq2}), we obtain
\begin{align}
\tilde{\nabla}^{\mu}\mathcal{T}_{\mu\nu}-R^{\lambda}_{~\rho\sigma\nu}\Delta_{\lambda}^{~\rho\sigma} = \delta(f)r^2_0\sin\theta_0\frac{8 l^6 + 4 l^4 r^2_0-r^6_0}{8l^4(l^2+r_0^2)^2}\neq 0
\end{align}
Therefore, there are no physical configuration of fluid and hyperfluid that will achieve this stargate configuration.
\subsection{Ideal stargate configuration: $T_{\mu\nu} = 0, \Delta_{\mu}^{~\nu\lambda} \neq 0$}\label{ayesat}
The ideal stargate configuration is achieved when
\begin{align}
\Gamma^{\mu}_{\nu\lambda} = \Gamma^{-\mu}_{\nu\lambda}
\end{align}
throughout the spacetime. This leads to vanishing of the Riemann tensor and commutativity of the covariant derivatives, which due to (\ref{identity}) implies
\begin{align}
\nabla_{\mu}\Delta_{\lambda}^{~\mu\nu} = 0
\end{align}
and the continuity equation is, therefore, trivially satisfied.
\section{Limiting behaviour of the physical stargate configuration}\label{limbeh}
We now look at the various limiting case of the stargate solution. For the limiting case of $\theta_0 = 0$, the (\ref{seT}) and (\ref{hypermomex}) gives
\begin{align}
T_{\mu\nu}\big|_{\theta_0 = 0} &= 0\\
\Delta^{~\mu\nu}_{\lambda}|_{\theta_0 = 0} &= 0
\end{align}
This is not surprising as this is supposed to be flat-space. For the limiting case of $\theta_0 = \pi$
\begin{align}
T_{\mu\nu}\big|_{\theta_0 = \pi} &= 0\\
\Delta^{~\mu\nu}_{\lambda}|_{\theta_0 = \pi} &= \Theta(\bar{f})2\frac{ r^2_0}{l}\sin\theta[\delta^{l}_{\lambda}(-\delta^{\mu}_{t}\delta^{\nu}_t+\delta^{\mu}_{l}\delta^{\nu}_{l})-\delta^{(\nu}_l\delta^{\mu)}_{\lambda}]\notag\\
&=2\frac{ r^2_0}{l}\sin\theta[\delta^{l}_{\lambda}(-\delta^{\mu}_{t}\delta^{\nu}_t+\delta^{\mu}_{l}\delta^{\nu}_{l})-\delta^{(\nu}_l\delta^{\mu)}_{\lambda}]
\end{align}
This is the full spherical Morris-Thorne wormhole supported entirely by a hypermomentum and has zero negative energy. This hypermomentum may be sourced by some hyperfluid which has been discussed in Section \ref{hypfluid}.
\section{Lack of conical singularities}\label{cs}
Consider the metric\footnote{where for $h(\theta) = \Theta(f)$, (\ref{sgm0}) can be recovered. For now, $h(\theta)$ is considered arbitrary.}
\begin{align}
ds^2 = -dt^2+dl^2 +[l^2+r^2_0 h(\theta)](d\theta^2+\sin^2\theta d\phi^2)
\end{align}
then one of the component of the Ricci tensor in metricity gravity is given by
\begin{align}
R_{l\theta} = \frac{l r^2_0 h'(\theta)}{[l^2+r^2_0 h(\theta)]^2}
\end{align}
This component is suspicious as we can see that by using
\begin{align}
&\lim_{z\to 0} z^{\Delta-d}\frac{z^{\Delta}}{(z^2+x_{\mu}x^{\mu})^{\Delta}} = \frac{\delta^{(d)}(x)}{C_{\Delta}} \notag\\
&C_{\Delta} = \frac{\Gamma(\Delta)}{\pi^{d/2}\Gamma(\Delta-\frac{d}{2})}\label{identity2}
\end{align}
where $d$ is the dimension of the coordinate $x^{\mu}$, as $l \to 0$
\begin{align}
R_{l\theta} \overset{l \to 0}{=} \frac{\pi r_0h'(\theta)}{l^2+r^2_0h(\theta)}\delta(\sqrt{h(\theta)})
\end{align}
there is a naive $\delta$-function singularity at the origin. Now, we make use of the identity
\begin{align}
\delta(f(x)) = \sum_{x_i \forall f(x_i) = 0} \frac{\delta(x-x_i)}{|f'(x_i)|}
\end{align}
so that we have
\begin{align}
R_{l\theta} \overset{l \to 0}{=} \frac{2\pi r_0h'(\theta)}{l^2+r^2_0h(\theta)}\left[ \sum_{\theta_i \forall h(\theta_i) = 0} \sqrt{h(\theta_i)}\frac{\delta(\theta-\theta_i)}{|h'(\theta_i)|}\right]
\end{align}
Now, if we take the overall factor inside the sum, we have
\begin{align}
R_{l\theta} \overset{l \to 0}{=} 2\pi\sum_{\theta_i \forall h(\theta_i) = 0}\frac{h'(\theta_i)}{|h'(\theta_i)|} \frac{r_0\sqrt{h(\theta_i)}}{l^2+r^2_0h(\theta_i)}\delta(\theta-\theta_i)
\end{align}
Again we make use of (\ref{identity}), since, $\sqrt{h(\theta_i)} = 0$. Since, $l^2$ is a $3$D norm, therefore,
\begin{align}
R_{l\theta} \overset{l \to 0}{=} -4\pi^3r^2_0\delta^{(3)}(l)\sum_{\theta_i \forall h(\theta_i) = 0}\frac{h'(\theta_i)}{|h'(\theta_i)|} |h(\theta_i)|\delta(\theta-\theta_i) = 0
\end{align}
Since, $h(\theta)$ was considered arbitrary, this must hold even for $h(\theta) = \Theta(f)$. The full Ricci tensor in the limit $l \to 0$ then becomes
\begin{align}
&R_{ll} \overset{l \to 0}{=} -\frac{2}{r^2_0h(\theta)} \quad R_{\theta\theta} \overset{l \to 0}{=} \frac{h(\theta)^{'2}-h(\theta)[\cot\theta h'(\theta)+h''(\theta)]}{2 h(\theta)^2} \quad R_{\phi\phi} \overset{l \to 0}{=} \sin^2\theta R_{\theta\theta}
\end{align}
It can be seen that a $\delta$-function singularity of the type $\delta(\theta_0-\theta)$ occurs in the above when $h(\theta) = \Theta(f)$. There are no additional $\delta$-function singularities at the origin.
\section*{Declarations}
\begin{itemize}
\item Funding: CSIR-UGC (JRF) (09/936(0212)/2019-EMR-I)
\item Conflict of interest: I, the sole author of this paper, declare that I have no conflicts of interest.
\item Ethics Approval: NA
\item Consent to participate: NA
\item Consent for publication: Yes
\item Availability of data, materials and code: NA
\item Author's Contributions: This is a single author paper. All sections have been worked out by the sole author of this paper.
\end{itemize}
\bibliographystyle{plain}
\bibliography{bibliyo}

\begin{thebibliography}{10}

\bibitem{bronnikov1973scalar}
KA~Bronnikov.
\newblock Scalar-tensor theory and scalar charge.
\newblock {\em Acta Phys. Pol}, page~B4, 1973.

\bibitem{clement1984class}
Gerard Clement.
\newblock A class of wormhole solutions to higher-dimensional general
  relativity.
\newblock {\em General relativity and gravitation}, 16(2):131--138, 1984.

\bibitem{einstein1935particle}
Albert Einstein and Nathan Rosen.
\newblock The particle problem in the general theory of relativity.
\newblock {\em Physical Review}, 48(1):73, 1935.

\bibitem{ellis1973ether}
Homer~G Ellis.
\newblock Ether flow through a drainhole: A particle model in general
  relativity.
\newblock {\em Journal of Mathematical Physics}, 14(1):104--118, 1973.

\bibitem{fisher1971product}
Brian Fisher.
\newblock The product of distributions.
\newblock {\em The Quarterly Journal of Mathematics}, 22(2):291--298, 1971.

\bibitem{fisher1992product}
Brian Fisher and Cheng Lin-Zhi.
\newblock The product of distributions on $r^m$.
\newblock {\em Comment. Math. Univ. Carolin}, 33(4):605--614, 1992.

\bibitem{garattini2020generalized}
Remo Garattini.
\newblock Generalized absurdly benign traversable wormholes powered by casimir
  energy.
\newblock {\em The European Physical Journal C}, 80(12):1--11, 2020.

\bibitem{Guendelman:2009er}
Eduardo~I Guendelman, Alexander Kaganovich, Emil Nissimov, and Svetlana
  Pacheva.
\newblock {Einstein-Rosen 'Bridge' Needs Lightlike Brane Source}.
\newblock {\em Phys. Lett. B}, 681:457--462, 2009.

\bibitem{iosifidis2020cosmological}
Damianos Iosifidis.
\newblock Cosmological hyperfluids, torsion and non-metricity.
\newblock {\em The European Physical Journal C}, 80(11):1--20, 2020.

\bibitem{iosifidis2020non}
Damianos Iosifidis.
\newblock Non-riemannian cosmology: The role of shear hypermomentum.
\newblock {\em International Journal of Geometric Methods in Modern Physics},
  18(2150129):1--19, 2021.

\bibitem{koh1992distributions}
EL~Koh and Li~Chen Kuan.
\newblock On the distributions $\delta^k$ and $(\delta')^ k$.
\newblock {\em Mathematische Nachrichten}, 157(1):243--248, 1992.

\bibitem{MR203392}
J.~Mikusi\'{n}ski.
\newblock On the square of the dirac delta-distribution.
\newblock {\em Bull. Acad. Polon. Sci. S\'{e}r. Sci. Math. Astronom. Phys.},
  14:511--513, 1966.

\bibitem{morris1988wormholes}
Michael~S Morris and Kip~S Thorne.
\newblock Wormholes in spacetime and their use for interstellar travel: A tool
  for teaching general relativity.
\newblock {\em American Journal of Physics}, 56(5):395--412, 1988.

\bibitem{Olmo:2015wwa}
Gonzalo~J. Olmo and D.~Rubiera-Garcia.
\newblock {Non-Riemannian geometry: towards new avenues for the physics of
  modified gravity}.
\newblock {\em J. Phys. Conf. Ser.}, 600(1):012041, 2015.

\bibitem{puetzfeld2008probing}
Dirk Puetzfeld and Yuri~N Obukhov.
\newblock Probing non-riemannian spacetime geometry.
\newblock {\em Physics Letters A}, 372(45):6711--6716, 2008.

\bibitem{van1959introduction}
JG~Van~der Corput.
\newblock Introduction to the neutrix calculus.
\newblock {\em Journal d’Analyse Math{\'e}matique}, 7(1):281--398, 1959.

\bibitem{woodward2011making}
James~F Woodward.
\newblock Making stargates: the physics of traversable absurdly benign
  wormholes.
\newblock {\em Physics Procedia}, 20:24--46, 2011.

\end{thebibliography}
\end{document}